\documentclass[english,aps,preprint]{revtex4}
\usepackage[T1]{fontenc}
\usepackage[latin1]{inputenc}
\usepackage{graphicx}
\usepackage{amssymb}

\makeatletter

\usepackage{graphicx}
\usepackage{graphicx}
\usepackage{graphicx}

\usepackage{babel}
\makeatother
\begin{document}

\title{Phase Space Tomography of Classical and Nonclassical Vibrational
States of Atoms in an Optical Lattice}

\author{J.F. Kanem, S. Maneshi, S.H. Myrskog, and A.M. Steinberg}

\affiliation{Centre for Quantum Information \& Quantum Control and Institute for
Optical Sciences, Department of Physics, University of Toronto, Canada }

\date{June 15, 2005 }


\begin{abstract}
Atoms trapped in optical lattice have long been a system of interest
in the AMO community, and in recent years much study has been devoted
to both short- and long-range coherence in this system, as well as
to its possible applications to quantum information processing. Here
we demonstrate for the first time complete determination of the quantum
phase space distributions for an ensemble of $^{85}Rb$ atoms in such
a lattice, including a negative Wigner function for atoms in an inverted
state. 
\end{abstract}
\maketitle

\section{Introduction}

Much of the excitement and promise of new fields such as quantum information
processing would not be possible without the development of sophisticated
techniques for manipulating and measuring quantum systems. In systems
such as ion traps, quantum dots, Bose-Einstein condensates, and entangled
photons, it is the fact that quantum states can be accurately prepared
and observed which has enabled a wide range of experimental advances,
and held out hope for the creation of practical quantum technologies.
In such systems, quantum states inevitably evolve into partially mixed
states, which can be characterized either by a density matrix, in
the case of a finite-dimensional Hilbert space, or by phase space
quasi-probability distributions such as the Wigner function\cite{wigner}
or the Husimi distribution\cite{husimi}. Techniques for extracting
these functions, generally referred to as {}``quantum tomography,''
have long been a topic of active research, in the context of the electromagnetic
field\cite{raymer,schiller}, Rydberg states\cite{walmsley}, neutral
atoms\cite{jessen1,mlynek1}, dissociating molecules\cite{molmer},
entangled photons\cite{white,kwiat,mitchell}, and ion traps\cite{cirac1,wineland}.

One system which has led to many interesting effects and proposals
is the {}``optical lattice,'' in which atoms are trapped in a periodic
potential formed by a standing wave of light beams\cite{prentiss,phillips1,jessen2}.
This analog condensed-matter system has been used to study Bragg scattering\cite{phillips2},
precision measurement using atom interferometry\cite{kasevich}, the
fragmentation of Bose condensates\cite{kasfrag,spekkens}, squeezed
states of atomic motion\cite{squeezed}, quantum feedback\cite{raithel},
the Mott insulator transition\cite{bloch1}, and quantum logic gates\cite{jessen3,cirac2,bloch2},
to name only a few examples. Typical probes of the system include
the Bragg scattering of probe beams, which is sensitive to the localization
$\left\langle X^{2}\right\rangle $ of the atoms; the time-varying
transmission of the lattice beams themselves, which is sensitive to
the instantaneous force exerted on the atoms, related to the centre-of-mass
position $\left\langle X\right\rangle $ in each well; and atom-interferometric
probes of long-range coherence. Here we present for the first time
a complete characterisation of the quantum state of atoms trapped
in optical-lattice wells, by extracting the Husimi and Wigner distributions
for atoms in several different initial states. In particular, we demonstrate
the {}``nonclassical'' Wigner function for atoms with a population
inversion, whose negative value at the origin is analogous to that
for the 1-photon Fock state and for the excited state of the single
ion in a trap.

\section{Theory}

The Husimi distribution is a well known quasi-probability distribution
related to a state's density matrix $\rho$ by the equation $Q\left(\alpha\right)=\frac{1}{\pi}\left\langle \alpha\right|\rho\left|\alpha\right\rangle $,
where $\left|\alpha\right\rangle $ is a coherent state, that is,
a Gaussian distribution centered at $\alpha$, with minimum-uncertainty
width which can be defined by a specified harmonic oscillator. It
has the practical advantage that its value for a particular $\alpha$
is an observable and can be obtained directly with one measurement.
Because it is always real and always positive it is sometimes referred
to as a classical quasi-probability distribution.

The Wigner distribution, which has a one-to-one correspondence with
the Husimi distribution, is particularly useful for identifying nonclassical
states such as Fock or inverted states, for which it takes on negative
value (as in fact it does for any non-Gaussian states). It can be
defined by the expression $W\left(x,p\right)=\frac{1}{\pi}\int_{-\infty}^{+\infty}\left\langle x+q\right|\rho\left|x-q\right\rangle \exp\left[-2ipq\right]dq$
\cite{leonhardt}. The Wigner distribution is unique in that it allows
one to determine the marginal probability distribution of either coordinate
by integrating over the other: $P\left(x\right)=\int W\left(x,p\right)dp$
and $P\left(p\right)=\int W\left(x,p\right)dx$.

The measurement of these quasi-probability distributions becomes particularly
simple in a harmonic oscillator. The Husimi distribution can be measured
by evaluating the expression\begin{equation}
Q\left(\left|\alpha\right|,\theta\right)=\frac{1}{\pi}\left\langle \alpha\right|\rho\left|\alpha\right\rangle =\frac{1}{\pi}\left\langle 0\right|D^{\dagger}\left(\left|\alpha\right|\right)R\left(\theta\right)\rho R^{\dagger}\left(\theta\right)D\left(\left|\alpha\right|\right)\left|0\right\rangle ,\eqnum{1}\end{equation}
 where $\theta=\arg\left\{ \alpha\right\} $. That is, instead of
projecting the unknown state $\rho$ onto a coherent state $\alpha$,
one can perform position displacement, $D\left(\left|\alpha\right|\right)$,
and rotation, $R\left(\theta\right)$, operations on the unknown state
and measure the overlap of the resulting state onto the ground state
$\left|0\right\rangle $, which is a straightforward process in our
experiment. Applying the displacement operator amounts to physically
displacing the state a distance of $x=2x_{0}\left|\alpha\right|$,
where $x_{0}=\left(\frac{\hbar}{2m\omega}\right)^{1/2}$ is the ground
state width of a particle of mass $m$ in a harmonic oscillator of
frequency $\omega$. In a harmonic oscillator $R\left(\theta\right)=\exp\left[-ia^{\dagger}a\theta\right]$
can be implemented by letting the state evolve under $\mathcal{H}=\hbar\omega\left(a^{\dagger}a+\frac{1}{2}\right)$
for a time $t=\theta/\omega$. How these operations are performed
experimentally and how the ground state population of a state is measured
will be described in section III.

The measurement of the Wigner distribution is simplified by recognizing
that for any symmetric non-degenerate potential and, in particular,
in a harmonic oscillator, $W\left(0,0\right)=\frac{1}{\pi}\int_{-\infty}^{+\infty}\left\langle q\right|\rho\left|-q\right\rangle dq=\frac{1}{\pi}\sum_{n=0}^{\infty}\left(-1\right)^{n}\left\langle n\right|\rho\left|n\right\rangle $.
For values away from the origin the displacement and rotation operators
can once again be utilized:\begin{equation}
W\left(x,p\right)=W\left(2x_{0}\Re e\left(\alpha\right),2p_{0}\Im m\left(\alpha\right)\right)=\frac{1}{\pi}\sum_{n=0}^{\infty}\left(-1\right)^{n}p\left(n\left|\alpha\right.\right),\eqnum{2}\end{equation}
 where $p_{0}=\sqrt{m\hbar\omega/2}$ and $p\left(n\left|\alpha\right.\right)=\left\langle n\right|D^{\dagger}\left(\left|\alpha\right|\right)R\left(\theta\right)\rho R^{\dagger}\left(\theta\right)D\left(\left|\alpha\right|\right)\left|n\right\rangle $
is the probability of finding the particle to be in state $n$ after
the application of the rotation and displacement operators to the
unknown state $\rho$. Therefore, the determination of the Wigner
distribution is reduced to performing population measurements after
applying displacement and rotation operators. This is similar to the
reconstruction of the Husimi distribution except that the final population
measurement requires the measurement of all states, ($\left|0\right\rangle ,\left|1\right\rangle ,\left|2\right\rangle ,...$),
not just the population of the ground state.

\section{Implementation}

\subsection{Optical Lattice Specifications}

A 1-D optical lattice is formed by the counter-propagating components
of two laser beams resulting in an intensity interference pattern
of the form $I\left(x\right)=I_{0}\cos^{2}\left(kx\sin\left(\frac{\gamma}{2}\right)\right)$,
where $k=\frac{2\pi}{\lambda}$ is the wave vector of the laser and
$\gamma$ is the angle of intersection of the two beams. This standing
light wave induces a light shift on the atoms resulting in a potential
$U\left(x\right)=I\left(x\right)\frac{\hbar\Gamma^{2}}{4\Delta I_{s}}$
where $\Delta$ is the detuning of the laser light from atomic resonance
and $\Gamma$and $I_{s}$ are the natural line-width and saturation
intensity of the atom respectively. Of importance to this experiment
is the fact that the individual wells of an optical lattice can be
approximated as harmonic oscillators. Therefore the theory of Husimi
and Wigner distribution measurements in harmonic oscillators can be
applied here. The oscillation frequency in each well $\omega$, and
lattice depth $U_{0}$, are related by the equation $\omega=\frac{4k_{L}}{\pi}\sqrt{\frac{U_{0}}{m}}$.
We find $\omega$ directly by measuring the period of Ramsey fringes
created by inducing oscillations with a spatial shift of the sinusoidal
potential\cite{myrskog}.

Our experiment starts with a cloud of $^{85}Rb$ atoms in a magneto
optical trap (MOT) which are then cooled in an optical molasses to
a temperature on the order of $10\mu K$. The optical lattice is turned
on during the MOT stage as cooling in the presence of the lattice
increases the lattice loading efficiency. The two laser beams intersect
at an angle of $\gamma=49.6^{\circ}$, resulting in a lattice vector
of $k_{L}=\frac{2\pi}{\lambda}\sin\left(\frac{\gamma}{2}\right)=\frac{\pi}{a}=3.38\cdot10^{6}m^{-1}$,
where $\lambda=780nm$ is the wavelength of the lattice light and
$a=0.93\mu m$ is the spatial period of the lattice. The lattice has
a detuning of $\Delta\simeq2\pi\cdot25GHz$ from the $F=3\Rightarrow F^{\prime}=4$
$D2$ trapping line of $^{85}Rb$ so as to make the scattering rate
negligible and to allow the sinusoidal lattice potential to be treated
as conservative. The lattice is tailored to support $2-4$ bound states,
depending on the experiment, which means a depth of $U_{0}\simeq10-40E_{r}$
where $E_{r}=\frac{\hbar^{2}k_{L}^{2}}{2m}$ is the scaled recoil
energy in the lattice direction. The lattice is formed in the vertical
direction so that atoms in unbound states will, in a time of $\lesssim10ms$,
leave the interaction region due to the pull of gravity. Each of the
lattice beams passes through an acousto-optic modulator (AOM). By
controlling the frequency and phase of the signal with which we drive
each AOM independently we are able to control the position, velocity
and acceleration of the lattice. Shifting the relative phase of the
lattice beams by $\phi$ displaces the lattice (or, in the rest frame
of the lattice, displaces the atoms) by a distance $d=a\phi/2\pi$.
Spatial shifts, used for the displacement operators $D\left(\left|\alpha\right|\right)$,
can be applied with a resolution on the order of $1nm$. Displacements
as large as twice the lattice spatial period occur in a time of $\lesssim0.5\mu s$.
With typical oscillation frequencies of $\omega/2\pi\simeq10^{4}Hz$,
this time interval for a shift can be considered to be instantaneous.

\subsection{State Preparation}

The preparation of the states whose quasi-probability distributions
we shall measure requires two steps. The first is to filter out the
ground state. By lowering the intensity of the lattice beams over
a time of $10ms$ until only one bound state is supported, and keeping
it there for $5ms$, all atoms in higher states become unbound and
fall out of the interaction region due to gravity. Afterwards we raise
the intensity back to its original level but are left with the ground
state in each well, with a typical contamination of the first excited
state of about $5-15\%$. A more detailed explanation of our ground
state preparation and measurement process can be found in reference
\cite{myrskog} and a similar technique is described in reference
\cite{mlynek2}. The implementation of the second part of the state
preparation depends on what state we wish to prepare.

In this paper we prepare and measure three states: a ground state,
a near coherent state and a state with a population inversion. For
the ground state no further action is necessary as it was prepared
in the filtering stage. A near coherent state, $\left|\beta\right\rangle $,
is prepared by shifting the potential by $\delta x=\frac{a}{6}=0.155\mu m$
(or a $60^{\circ}$ phase shift). The magnitude of $\beta$ is related
to the displacement by $\left|\beta\right|=\delta x\left(\frac{m\omega}{2\hbar}\right)^{1/2}=0.88$.
For both the ground and coherent state a lattice depth of $37E_{r}$
was chosen, supporting $4$ bound states with an oscillation frequency
of $\omega=48.33kHz$. The deviation of the near coherent state which
we prepare from an actual coherent state is due to the finite depth
of the lattice. With only $4$ bound states only the first $4$ terms
of the coherent state are present in the lattice. This does not significantly
detract from the validity of the approximation, as the amplitude of
the $n^{\textrm{th}}$ term in a harmonic oscillator coherent state
is $\exp\left[-\left|\beta\right|^{2}/2\right]\frac{1}{\sqrt{n!}}\left|\beta\right|^{n}$,
which when $|\beta|=0.88$ is quite small for $n\geq4$. This state
was allowed to rotate in the potential for a time $t=20\mu s$, giving
it a rotation of $\theta=0.97$ radians.

Creation of an inverted state begins with the preparation of the ground
state, this time in a lattice containing only two bound states ($\omega=32.2kHz$).
Next, the potential is given a $60^{\circ}$ phase shift ($0.155\mu m$),
held there for $80\mu s$ and then shifted back to its original position.
We have found that this process excites a large number of atoms into
higher states\cite{maneshi}. After waiting several milliseconds in
order to let unbound atoms leave the interaction region we are left
with what will be shown to be an incoherent mixture of ground and
first-excited-state atoms in a ratio of roughly $3$ to $7$.

\subsection{State Measurement}

Here we describe the process used to determine the Husimi distribution
for the ground state and coherent state. As per equation 1 the measurement
of $Q\left(\left|\alpha\right|,\theta\right)$ takes three steps.
First, we allow the atoms to undergo free evolution for a time $t$
to let the state oscillate in the harmonic-like potential for a rotation
of $\theta=\omega t$ in phase space. We use a total of $27$ angles
(or wait times, $t$) separated by $0.24$ radians (or $5\mu s$),
spanning a range of $\theta\epsilon\left[0,2\pi\right]$ ($\delta t\epsilon\left[0,130\right]\mu s$).
Second, a displacement of $x=2x_{0}\left|\alpha\right|$ is applied
to the lattice. This is the displacement operator, $D\left(\left|\alpha\right|\right)$.
A total of $19$ different displacements are used, each separated
by $25.8nm$ or a $10^{\circ}$ phase shift of the potential for a
total range of $x\epsilon\left[0,465\right]nm$ or $\left[0,180\right]$
degrees. Along with the phase component, the total number of measurements
is then $513$ not including repeated measurements for statistical
analysis and averaging. Next, a projection of this new state onto
the ground state is performed. This first requires the same process
that was used to filter the ground state during the state preparation:
the lattice laser intensity is lowered until only the ground state
remains. After waiting $\sim20ms$ in order to let the unbound atoms
become spatially separated from those still bound the atoms are illuminated
with resonant light, and their fluorescence collected in a CCD camera.
With this image the relative population of ground state atoms, as
a fraction of total atom number, is measured.

Figure 1 is a phase space diagram of the Husimi distribution for the
coherent state. As expected, it is a Gaussian displaced from the origin
by $\left|\beta\right|=0.88$ at an angle of roughly $55^{\circ}$.
Since a coherent state is merely a ground state displaced from the
origin in phase space both distributions should have the same width.
Figure 2 shows a cross section of the Husimi distribution for the
ground state along with one of the coherent state. Fits of these Gaussians
give $rms$ widths of $144nm$ and $147nm$ and peaks of $0.267$
and $0.257$ for the ground and coherent states respectively. Husimi
distributions, by definition, are normalized. By integrating the curve
shown in figure 1 we find from$\int Q\left(\left|\alpha\right|,\theta\right)\left|\alpha\right|d\left|\alpha\right|d\theta=\int Q\left(\frac{r}{2x_{rms}},\theta\right)\frac{r}{4x_{rms}^{2}}drd\theta=1$
that $x_{rms}=96.3nm$. Here we find a surprising result. The ground
state width of a harmonic oscillator of this frequency is $x_{0}=\left(\frac{\hbar}{2m\omega}\right)^{1/2}=88.0nm$
but we find $x_{rms}=96.3nm=1.09\cdot x_{0}$. We believe that this
discrepancy is due to inhomogeneities in the intensity of our lattice
beams. We have separately measured the dephasing time $T_{2}$ to
be on the order of 2 or 3 oscillation periods in our lattice\cite{myrskog,maneshi},
implying $\Delta\omega\approx0.4\omega$. As $x_{0}^{2}=\hbar/2m\omega$,
we expect the rms width of our atom clouds to be $x_{rms}=(\hbar/2m)\langle1/\omega\rangle=x_{0}^{2}[1+(\Delta\omega/\omega)^{2}+(\Delta\omega/\omega)^{4}+\ldots)]$.
For $\Delta\omega/\omega\sim0.4$, this means $x_{rms}\approx1.08x_{0}$,
consistent with our observations.

In addition, the peak of the Husimi distribution of a true ground
state is $\frac{1}{\pi}$. Our measured distribution has a peak value
of $0.267=0.84/\pi$, obtained from a direct measurement of the ground-state
population to be $84\%$. A harmonic-oscillator state with $16\%$
population in the first excited state, when the ground-state width
is taken to be $x_{rms}$ as calculated above, has a Husimi distribution
with a width of $141$ nm, consistent with our measured value.

Following equation 2 the Wigner distribution for the inverted state
is measured by first applying the rotation and displacement operators
$R\left(\theta\right)$ and $D\left(\left|\alpha\right|\right)$,
as in the Husimi reconstruction. For the rotation operators we sample
$41$ different rotation angles separated by $0.16$ radians spanning
a range of $\theta\epsilon\left[0,2\pi\right]$ equivalent to waiting
times in the range $\delta t\epsilon\left[0,200\right]\mu s$ with
a resolution of $5\mu s$. We again use $19$ different spatial shifts
for the displacement operators with a range of $x\epsilon\left[0,465\right]nm$
giving a total number of $779$ measurements for the Wigner measurement.

After the rotation and displacement operations are implemented, the
state populations must be measured. Since the lattice holds two bound
states any atoms excited to higher states by the rotation and displacement
operators become unbound. After waiting $5ms$ in order to let these
atoms become spatially separated we once again lower the lattice laser
intensity until only one bound state remains. After waiting another
$20ms$ in order to let these first-excited-state atoms leave the
interaction region we then capture a fluorescence image. From this
image we determine the ground and first-excited-state populations
as well as how many atoms were in states $n\geq2$. Therefore we are
able to construct the Wigner distribution with the following caveat:
measurements of $W\left(x,p\right)$ become increasingly uncertain
for values for which the displacement operator $D\left(\left|\alpha\right|\right)$
excites more atoms into $n\geq2$, since only the first two terms
of equation 2 are known exactly.

Figure 3 shows a cross section of our best estimate of the Wigner
distribution: $W^{\prime}\left(x,0\right)\equiv\frac{1}{\pi}\left(p\left(0\left|\alpha\right.\right)-p\left(1\left|\alpha\right.\right)\right)$.
At the origin we see that the distribution is negative, a nonclassical
signature which is characteristic of a population inversion. The inset
shows the absolute upper and lower bounds, based on our data, of the
Wigner distribution due to our lack of knowledge about the populations
of states higher than $n=1$. The upper bound is obtained by assuming
that all atoms lost during the displacement and rotation operators
were in the 2nd excited state; $\frac{1}{\pi}\left(p\left(0\left|\alpha\right.\right)-p\left(1\left|\alpha\right.\right)+p\left(n>1\left|\alpha\right.\right)\right)$,
while the lower is constructed by assuming they were in the 3rd; $\frac{1}{\pi}\left(p\left(0\left|\alpha\right.\right)-p\left(1\left|\alpha\right.\right)-p\left(n>1\left|\alpha\right.\right)\right)$.
Of course neither one of these extremes can be the case as they are
not physical. It is known that the Wigner distribution for any state
must go to zero as $x,p\rightarrow\infty$. In addition, one can get
a sense of the magnitude of the disagreement between $W$ and $W^{\prime}$
by investigating the normalisation of the three curves; $\int W^{\prime}\left(x,p\right)dxdp=0.42$,
which is clearly not consistent with $1$. The same normalisation
integral, however, gives values of $10.32$ and $-9.49$ for the upper-
and lower-bound curves, respectively. Clearly, the real Wigner function
lies between $W^{\prime}$ and the upper bound, but much closer to
the former. Theoretically, one expects a significant additional contribution
from the second excited state for $|\alpha|\sim1$, but for larger
values of $|\alpha|$ this should be cancelled out by higher populations
in the third excited state. Near the origin, there was no population
in states with $n\geq2$, so there is no uncertainty.

Figure 4 shows a 3-D representation of the Wigner distribution measurement.
Note the rotational symmetry, which indicates a lack of coherence
between energy eigenstates. Usually processes such as the displacement
operator create coherent superpositions of states such as in the case
with our presented Husimi distribution. The lack of any coherence
properties here is due to the combination of dephasing processes inherent
in our optical lattice\cite{myrskog} and the long delay between the
population inversion creation and the time of measurement. Therefore
the result of our state creation, as shown by figures 3 and 4 is an
incoherent mixture of ground and 1st excited state atoms with most
atoms being in the latter, as evidenced by the negative component
of the Wigner distribution.

\section{Conclusion}

Making use of a newly developed technique to measure state populations
of the vibrational states of atoms in 1-$\mu m$ lattice wells, and
our ability to perform arbitrary translations in phase space, we have
reconstructed Husimi and Wigner phase space distributions for atoms
in ground, coherent, and inverted states of oscillation in an optical
lattice. This is the first complete quantum characterisation of the
state of motion of atoms in such a system. The ground and coherent
states are largely consistent with expectations, possessing essentially
the same shape and width, although they indicate both some admixture
of the excited state, which is understandable in light of our preparation
procedure, and an apparent underestimate of the width of the ground-state
wave function relative to the experimental measurement, which is probably
due to inhomogeneities in the lattice beams. We observe a nonclassical
signature in the Wigner function, reaching a maximum negative value
of $-0.12$ at the origin, in the case of the inverted distribution.

\section{Acknowledgments}

We would like to thank Matt Partlow for helpful discussions and assistance
with the manuscript, and John Sipe for many informative conversations
about Wigner distributions. We acknowledge financial support from
NSERC, from the CIAR, and from the DARPA QuIST program (managed by
the AFOSR under agreement no. F49620-01-1-0468).

 \newpage

\begin{center}\includegraphics[%
  width=7cm]{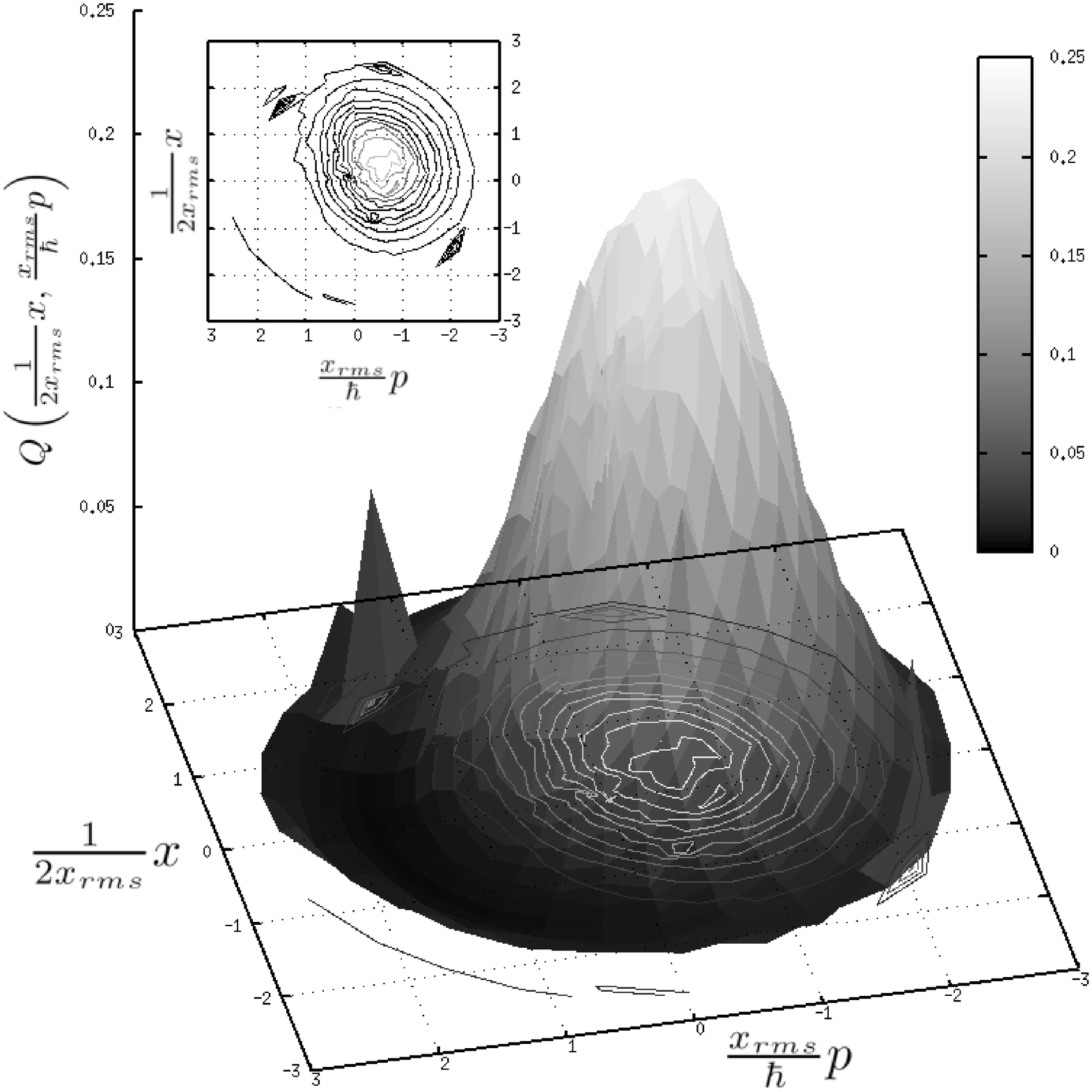}\end{center}

Figure 1: Phase space representation of the Husimi distribution of
a coherent state $\left|\left|\beta\right|,\theta\right\rangle $
for $\left|\beta\right|=0.88$, $\theta=0.97$. From the contour plot
in the inset one can see the displacement of the Gaussian from the
origin. Axis units are in $x/2x_{rms}$ where $x_{rms}=96.3nm$.

\begin{center}\includegraphics[%
  width=7cm]{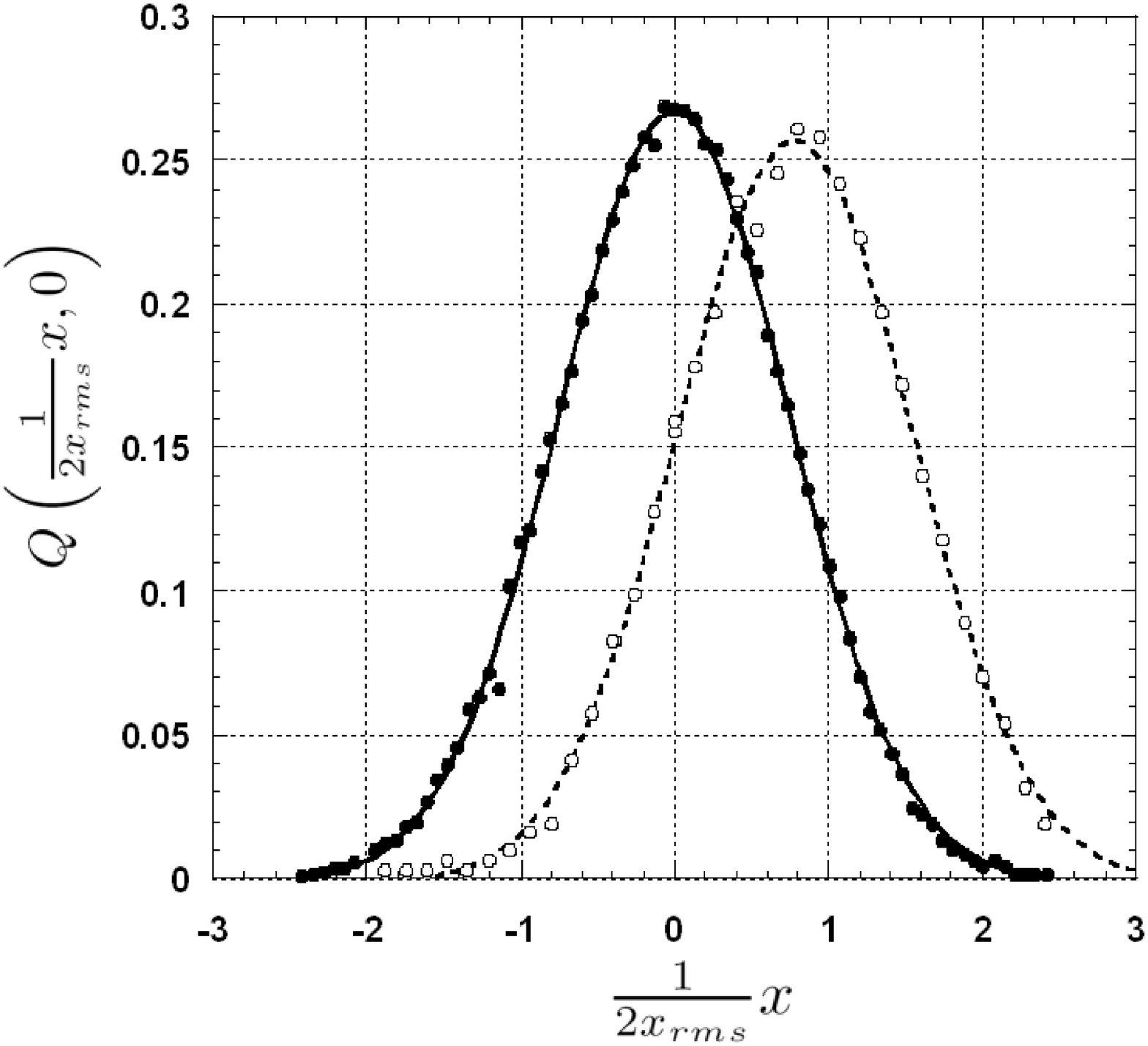}\end{center}

Figure 2: Cross sections of the Husimi distributions of the ground
state (closed circles) and coherent state (open circles). As expected
they have virtually the same width but the coherent state is displaced
from the origin. In addition, the height and width, when compared
to the true ground state of a harmonic oscillator of this frequency,
show that there is some contamination of the state by first excited
state atoms.

\begin{center}\includegraphics[%
  width=7cm]{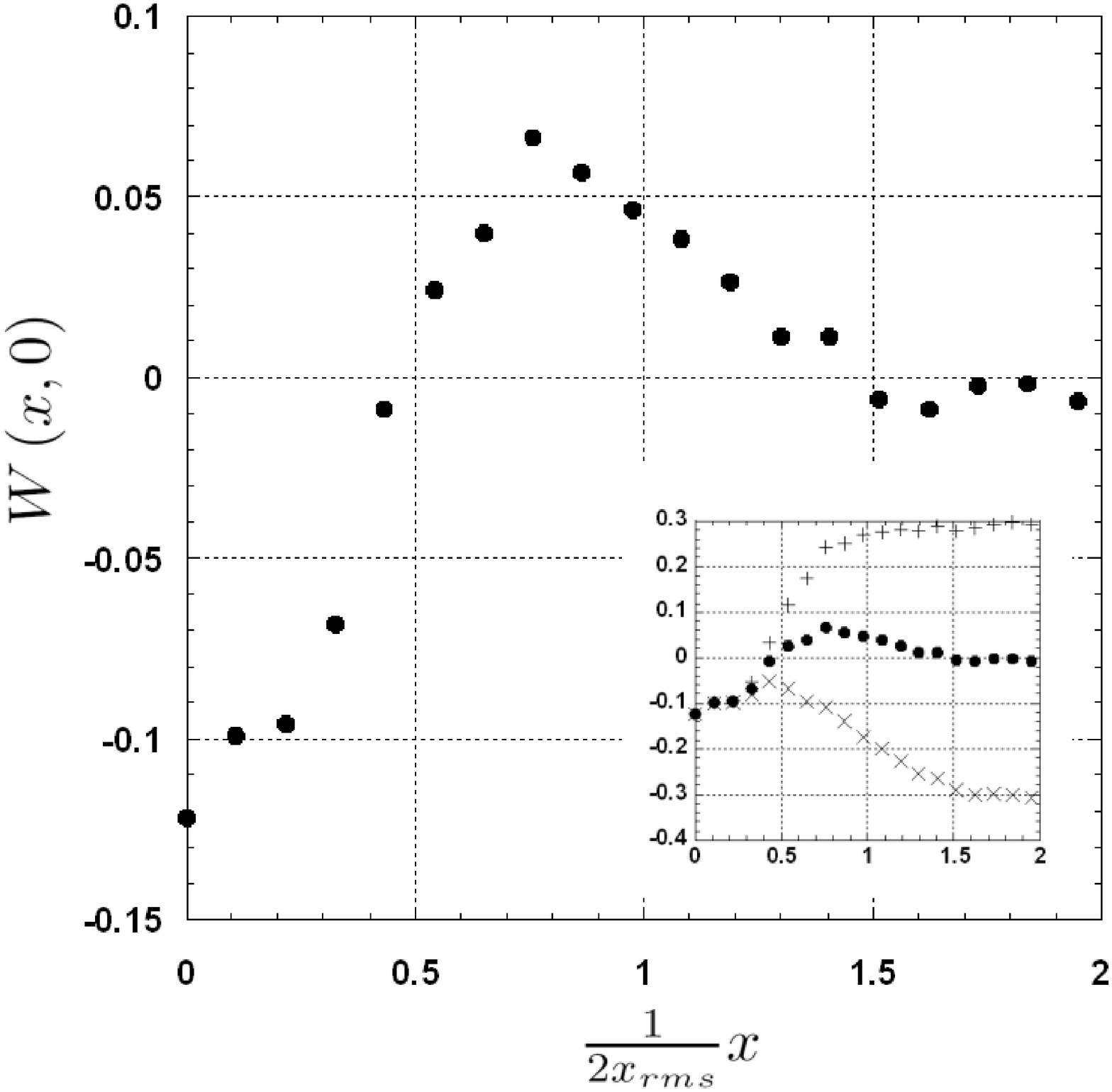}\end{center}

Figure 3: Cross section of the measured Wigner distribution. Negative
values at the origin are indicative of a population inversion. Inset
shows absolute upper (+) and lower ($\times$) bounds of the distribution
as explained in the text. Axis units are in $x/2x_{rms}$ where $x_{rms}=119.4nm$

\begin{center}\includegraphics[%
  width=7cm]{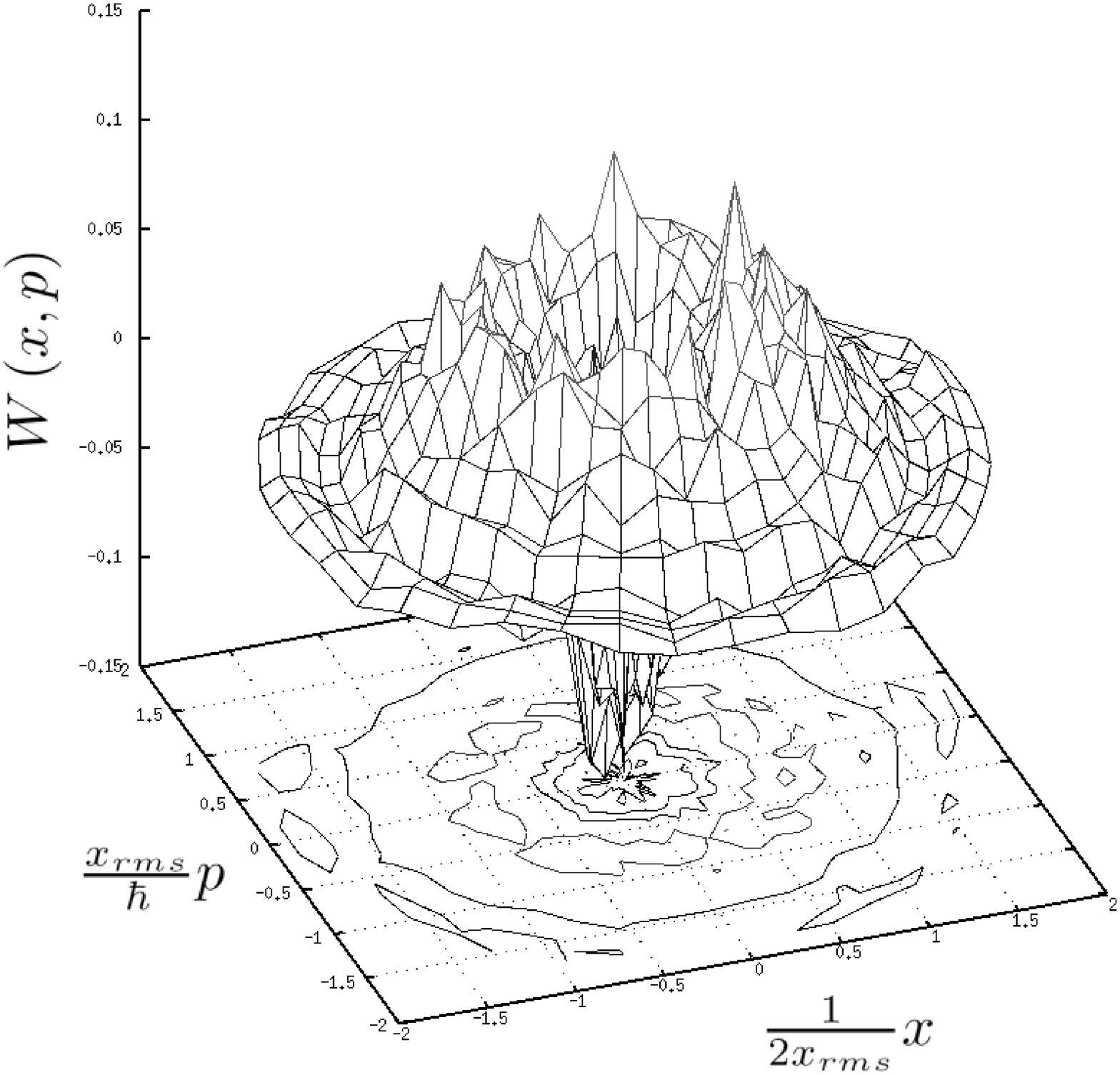}\end{center}

Figure 4: 3-D image of the measured Wigner distribution. Of note is
the cylindrical symmetry showing a lack of coherence between the component
energy eigenstates. 
\end{document}